\documentclass[prb,aps,twocolumn,floatfix,psfig,epsf]{revtex4}
\usepackage{amssymb}
\usepackage{graphicx}

\input{tcilatex}

\begin{document}

\title{The quantum critical behavior of antiferromagnetic itinerant systems
with van Hove singularities of electronic spectrum}
\author{A. Katanin}

\affiliation{
Institute of Metal Physics, Kovalevskaya str. 18, 620041,
Ekaterinburg, Russia\\
Max-Planck-Institut f\"ur Festk\"orperforschung, 70569 Stuttgart, Germany}

\begin{abstract}
The interplay of magnetic and superconducting fluctuations in two
dimensional systems with van Hove singularities in the electronic spectrum
is considered within the functional renormalization group (fRG) approach.
While the fRG flow has to be stoped at a certain minimal temperature $T_{%
\text{RG}}^{\min }$, we study temperature dependence of magnetic and
superconducting susceptibilities both, above and below $T_{\text{RG}}^{\min
} $, which allows to obtain the resulting ground state phase diagram. Close
to half filling the fRG approach yields two quantum phase transitions: from
commensurate antiferromagnetic to incommensurate phase and from the
incommensurate to paramagnetic phase, the region of the commensurate
magnetic phase is possibly phase separated away from half filling. Similarly
to results of Hertz-Moriya-Millis approach, the temperature dependence of
the inverse (incommensurate) magnetic susceptibility at the quantum phase
transition from incommensurate to paramagnetic phase is found almost linear
in temperature.
\end{abstract}

\maketitle

\section{Introduction}

The quantum critical points (QCP) in itinerant magnets have being
investigated during long time. Moriya theory \cite{Moriya} was first attempt
to describe thermodynamic properties near QCP. This theory was further
developed within Hertz-Millis renormalization group approach\cite{HM}. In
more than two dimensions Hertz-Moriya-Millis approach predicts that the
magnetic transition temperature $T_{c}$ depends on the distance $\delta $ to
the QCP as $T_{c}\sim \delta ^{z/(d+z-2)}$. In two-dimensional systems, the
long-range magnetic order at finite temperatures is prohibited according to
the Mermin-Wagner theorem, but the quantum phase transition is accompanied
by vanishing of the temperature of the crossover to the regime with
exponentially large correlation length (renormalized-classical regime), $%
T^{\ast }\sim \delta $.

The applicability of Hertz-Moriya-Millis (HMM) approach to magnetic systems
was however recently questioned because of expected strong momenta- and
frequency dependence of the paramagnon interaction vertices\cite{Chub} and
possible non-analytical dependence of the magnetic susceptibility\cite{Non}
which arises due to strong electron-paramagnon interaction. Studying the
problem of quantum critical behavior of these systems in terms of fermionic
degrees of freedom may be helpful to obtain concentration dependence of the
crossover temperature and provide valuable information about their magnetic
properties near quantum critical points.

Itinerant systems with van Hove singularities in the electronic spectrum
have strong momentum dependence of interaction vertices due to peculiarity
of the electronic dispersion, and, therefore represent an interesting
example for studying the quantum critical behavior. The competition of
different kinds of fluctuations, and even long range orders is important in
the presence of van Hove singularities, which makes formulation of effective
boson-fermion theories rather complicated. At the same time, fermionic
approaches can treat naturally both, (anti)ferromagnetic and superconducting
fluctuations, which were considered to be important near magnetic quantum
phase transitions in systems with van Hove singularities in the electronic
spectrum.

The simplest mean-field analysis of the Hubbard model is insufficient to
study quantum critical behavior; due to locality of the Coulomb repulsion in
this model it is also unable to investigate the range of existence of
unconventional (e.g., d- or p-wave) superconducting order, and introduction
of the nearest-neighbor interaction is required in this approach\cite{MF}.
To study the competition of magnetism and superconductivity in the Hubbard
model, more sophisticated approaches, e.g. cluster methods\cite%
{Licht,Jarrell} and functional renormalization group (fRG) approaches\cite%
{Metzner1,SalmHon,SalmHon1,KK} were used. The fRG approaches are not limited
by the system (cluster) size and offer a possibility to study both, magnetic
and superconducting fluctuations, as well as their interplay at weak and
intermediate coupling.

The fRG approaches were initially applied to the paramagnetic
non-superconducting (symmetric) phase to study the dominant type of
fluctuations in different regions of the phase diagram\cite%
{Metzner1,SalmHon,SalmHon1,KK}. Although these approaches suffered from the
divergence of vertices and susceptibilities at low enough temperatures near
the magnetic or superconducting instabilities, comparing susceptibilities
with respect to different types of order at the lowest accessible
temperature provided a possibility to deduce instabilities in different
regions of the phase diagram. In fact, the temperature where the vertices
and susceptibilities diverge in the one-loop approach, is related to above
discussed temperature $T^{\ast }$ of the crossover to the `strong-coupling'
regime with exponentially large magnetic correlation length.

To access the region $T<T^{\ast },$ the combination of the fRG and
mean-field approach was proposed in Ref. \cite{Metzner}, which was also able
to study possible coexistence of magnetic and superconducting order at $T=0$
(the magnetic order parameter was assumed to be commensurate). More
sophisticated fRG approach in the symmetry-broken phase \cite{MetznerfRGSB}
was developed recently to avoid application of the mean-field approach after
the RG flow; the application of this method was however so far restricted by
the attractive Hubbard model, because of complicated structure of the
resulting renormalization group equations.

So far only susceptibilities corresponding to spin and charge fluctuations
with commensurate wavevectors, as well, as to superconducting fluctuations
were carefully investigated. In the present paper we use the fRG approach in
the symmetric phase\cite{SalmHon1,KK} and perform an accurate analysis of
temperature dependence of susceptibilities with respect to both,
commensurate and incommensurate magnetic order, as well as superconducting
order. We propose extrapolation method which allows us to study
thermodynamic properties both above and below the temperature at which the
fRG flow is stopped, and extract the crossover temperature $T^{\ast }$. This
gives us a possibility to obtain phase diagram, capturing substantial part
of the fluctuations of magnetic and superconducting order parameters without
introducing symmetry breaking. Contrary to the functional renormalization
group analysis in the symmetry broken phase\cite{MetznerfRGSB}, the
presented method can be easily generalized to study instabilities with
different type of the order parameters.

\section{Method}

We consider the 2D $t$-$t^{\prime }$ Hubbard model $H_{\mu }=H-(\mu
-4t^{\prime })N$ with 
\begin{equation}
H=-\sum_{ij\sigma }t_{ij}c_{i\sigma }^{\dagger }c_{j\sigma
}+U\sum_{i}n_{i\uparrow }n_{i\downarrow }\,,  \label{H}
\end{equation}%
where $t_{ij}=t$ for nearest neighbor (nn) sites $i$, $j$, and $%
t_{ij}=-t^{\prime }$ for next-nn sites ($t,t^{\prime }>0$) on a square
lattice; for convenience we have shifted the chemical potential $\mu $ by $%
4t^{\prime }$. We employ the fRG approach for one-particle irreducible
generating functional and choose temperature as a natural cutoff parameter
as proposed in Ref. \cite{SalmHon1}. This choice of cutoff allows us to
account for excitations with momenta far from and close to the Fermi
surface. Neglecting the frequency dependence of interaction vertices, the RG
differential equation for the interaction vertex $V_{T}\equiv $ $V(\mathbf{k}%
_{1},\mathbf{k}_{2},\mathbf{k}_{3},\mathbf{k}_{4})$ has the form \cite%
{SalmHon1} 
\begin{equation}
\frac{\mathrm{d}V_{T}}{\mathrm{d}T}=-V_{T}\circ \frac{\mathrm{d}L_{\mathrm{pp%
}}}{\mathrm{d}T}\circ V_{T}+V_{T}\circ \frac{\mathrm{d}L_{\mathrm{ph}}}{%
\mathrm{d}T}\circ V_{T}\,,  \label{dV}
\end{equation}%
where $\circ $ is a short notation for summations over intermediate momenta
and spin, momenta $\mathbf{k}_{i}$ are supposed to fulfill the momentum
conservation law $\mathbf{k}_{1}+\mathbf{k}_{2}=\mathbf{k}_{3}+\mathbf{k}%
_{4},$%
\begin{equation}
L_{\text{ph,pp}}(\mathbf{k},\mathbf{k}^{\prime })=\frac{f_{T}(\varepsilon _{%
\mathbf{k}})-f_{T}(\pm \varepsilon _{\mathbf{k}^{\prime }})}{\varepsilon _{%
\mathbf{k}}\mp \varepsilon _{\mathbf{k}^{\prime }}},  \label{Lphpp}
\end{equation}%
and $f_{T}(\varepsilon )$ is the Fermi function. The upper signs in Eq. (\ref%
{Lphpp}) stand for the particle-hole ($L_{\text{ph}}$) and the lower signs
for the particle-particle ($L_{\text{pp}}$) bubbles, respectively. Eq. (\ref%
{dV}) is solved with the initial condition $V_{T_{0}}(\mathbf{k}_{1},\mathbf{%
k}_{2},\mathbf{k}_{3},\mathbf{k}_{4})=U$; the initial temperature is chosen
as large as $T_{0}=10^{3}t$. The evolution of the vertices with decreasing
temperature determines the temperature dependence of the susceptibilities
according to \cite{SalmHon1} 
\begin{eqnarray}
\displaystyle{\frac{\mathrm{d}\chi _{m}}{\mathrm{d}T}} &=&\sum_{\mathbf{k}%
^{\prime }}\mathcal{R}_{\mathbf{k}^{\prime }}^{m}\mathcal{R}_{\pm \mathbf{k}%
^{\prime }+\mathbf{q}_{m}}^{m}\displaystyle{\frac{\mathrm{d}L_{\text{ph,pp}}(%
\mathbf{k}^{\prime },\pm \mathbf{k}^{\prime }+\mathbf{q}_{m})}{\mathrm{d}T}},
\label{dH} \\
\displaystyle{\frac{\mathrm{d}\mathcal{R}_{\mathbf{k}}^{m}}{\mathrm{d}T}}
&=&\mp \sum_{\mathbf{k}^{\prime }}\mathcal{R}_{\mathbf{k}^{\prime
}}^{m}\Gamma _{m}^{T}(\mathbf{k},\mathbf{k}^{\prime })\displaystyle{\frac{%
\mathrm{d}L_{\text{ph,pp}}(\mathbf{k}^{\prime },\pm \mathbf{k}^{\prime }+%
\mathbf{q}_{m})}{\mathrm{d}T}}.  \nonumber
\end{eqnarray}%
Here the three-point vertices $\mathcal{R}_{\mathbf{k}}^{m}$ describe the
propagation of an electron in a static external field, $m$ denotes one of
the instabilities: antiferromagnetic (AF) with $\mathbf{q}_{m}=(\pi ,\pi ),$
incommensurate magnetic (\textbf{Q}) with the wave vector $\mathbf{q}_{m}=%
\mathbf{Q}$, or d-wave superconducting (dSC)\ with $\mathbf{q}_{m}=0$ (upper
signs and ph correspond to the magnetic instabilities, lower signs and pp to
the superconducting instability);%
\begin{equation}
\Gamma _{m}^{T}(\mathbf{k},\mathbf{k}^{\prime })=\left\{ 
\begin{array}{cl}
V_{T}(\mathbf{k},\mathbf{k}^{\prime },\mathbf{k}^{\prime }+\mathbf{q}_{m}) & 
m=\text{AF or \textbf{Q},} \\ 
V_{T}(\mathbf{k},-\mathbf{k+q}_{m},\mathbf{k}^{\prime }) & m=\text{dSC.}%
\end{array}%
\right.  \label{Gamma}
\end{equation}

The initial conditions at $T_{0}$ for Eqs. (\ref{dH}) are $\mathcal{R}_{%
\mathbf{k}}^{m}=f_{\mathbf{k}}$ and $\chi _{m}=0$, where the function $f_{%
\mathbf{k}}$ belongs to one of the irreducible representations of the point
group of the square lattice, e.g. $f_{\mathbf{k}}=1$ for the magnetic
instabilities and $f_{\mathbf{k}}=(\cos k_{x}-\cos k_{y})/A$ for the d-wave
superconducting instability, with a normalization coefficient $A=(1/N)\sum_{%
\mathbf{k}}f_{\mathbf{k}}^{2}$. 
\begin{figure}[tb]
\includegraphics[width=8.7cm]{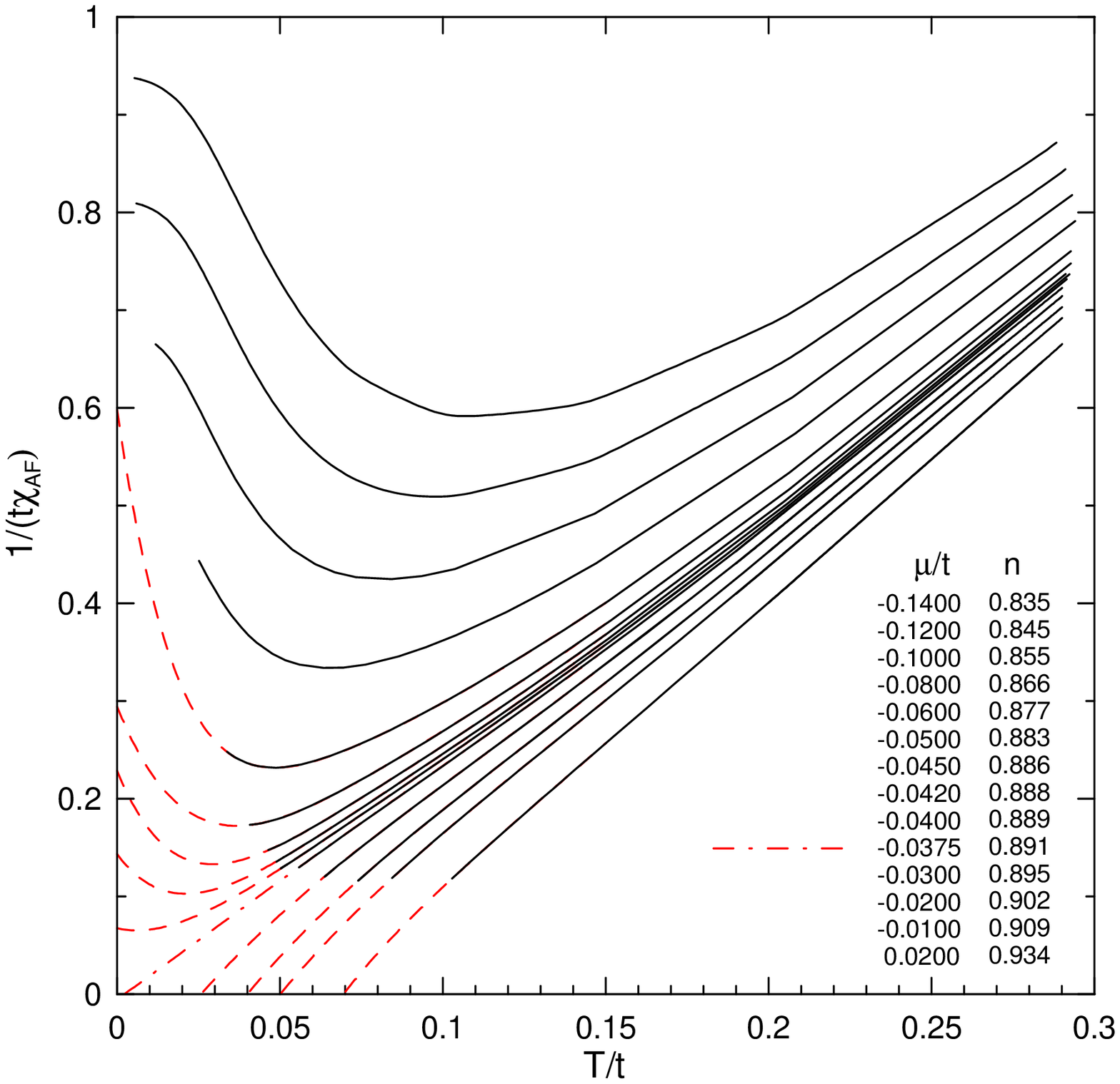} \vspace{-3mm}
\caption{(Color online) Temperature dependences of the inverse
antiferromagnetic susceptibility at $t^{\prime }/t=0.1t$, $U=2.5t$, and
different values of the chemical potential (the list of the chemical
potentials and fillings corresponds to the curves from top to bottom, the
smallest $\protect\mu $ corresponds to upper curve). Dashed lines show the
extrapolation of the inverse susceptibilities to the temperature region $%
T<T_{\text{RG}}^{\mathrm{min}}$ by polynomials of 6-th order}
\label{fig1}
\end{figure}
\begin{figure}[tb]
\includegraphics[width=8.7cm]{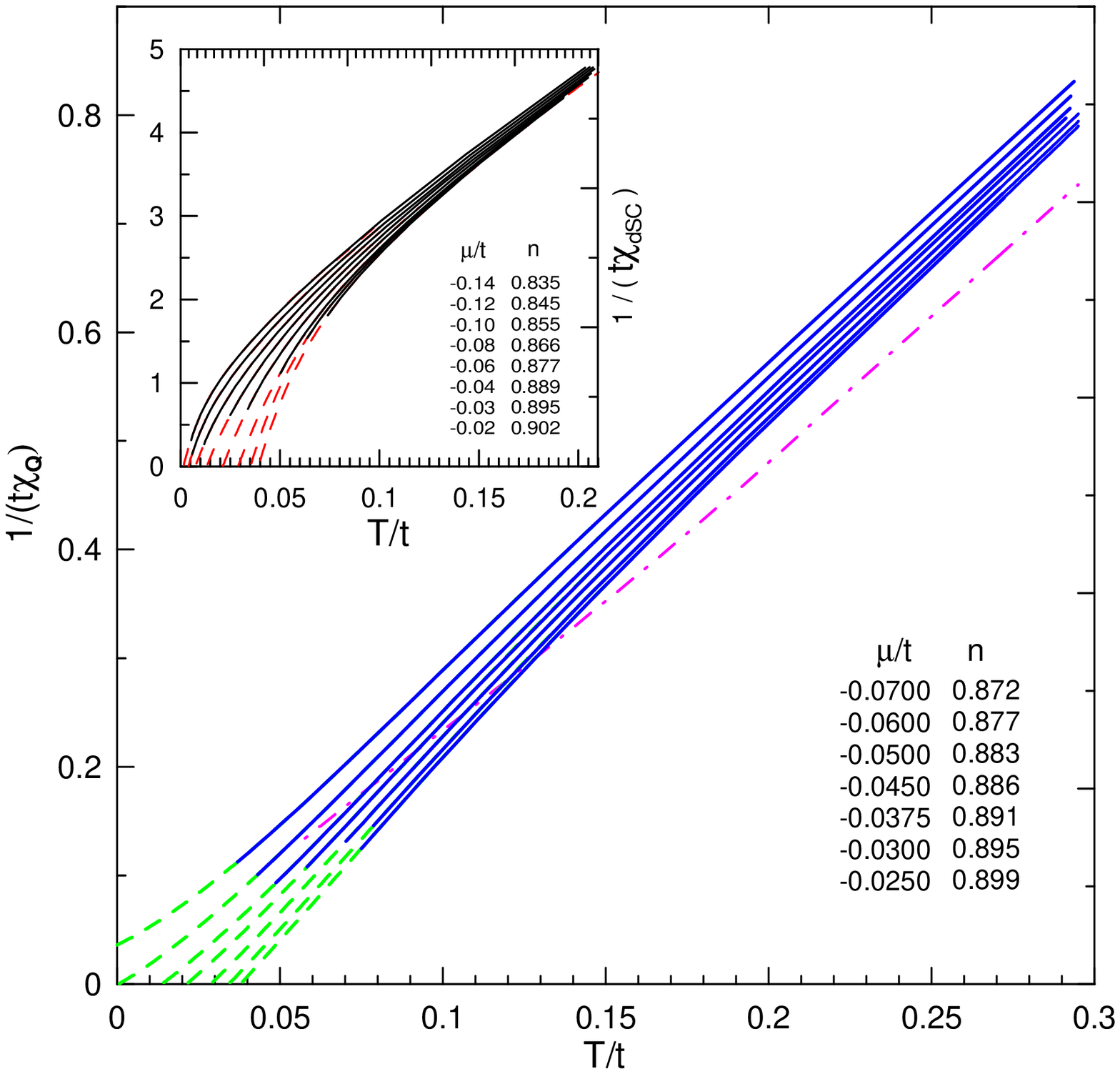} \vspace{-3mm}
\caption{(Color online) Temperature dependences of the inverse magnetic
susceptibility at $t^{\prime }/t=0.1t$, $U=2.5t$ and the incommensurate
wavevector determined by a maximum $T_{c}^{\text{\textbf{Q}}}$, dashed lines
show the extrapolation to $T<T_{\text{RG}}^{\mathrm{min}}$. Dot-dashed line
shows the inverse commensurate susceptibility at $\protect\mu =\protect\mu %
_{c}^{\mathrm{AF}}\approx -0.0375t$. The inset shows temperature dependences
of the inverse susceptibility with respect to d-wave superconducting pairing
at different values of the chemical potential}
\label{fig2}
\end{figure}
To solve the Eqs. (\ref{dV}) and (\ref{dH}) we discretize the momentum space
in $N_{p}=48$ patches using the same patching scheme as in Ref. \cite%
{SalmHon1}. This reduces the integro-differential equations (\ref{dV}) and (%
\ref{dH}) to a set of 5824 differential equations, which were solved
numerically. In the present paper we perform the renormalization group
analysis down to the temperature $T_{\text{RG}}^{\min },$ at which vertices
reach some maximal value (we choose $V_{\max }=18t$).

To obtain the behavior of the susceptibilities at $T<T_{\text{RG}}^{\min }$
we extrapolate obtained temperature dependence of the inverse
susceptibilities by fitting this dependence above (but close to) $T_{\text{RG%
}}^{\min }$ by polynomials of $5$-th to $7$-th order. We identify the
crossover temperature $T_{m}^{\ast }$ to the regime of strong correlations
of the order parameter denoted by $m$ from the condition that the
extrapolated $\chi _{m}^{-1}(T_{m}^{\ast })=0$ (we assume that the
susceptibilities are almost analytic functions of temperature in the
crossover regime). We have checked that the obtained $T_{m}^{\ast }$
essentially depends on neither the order of polynomial, used for the
fitting, nor on the fitting range. Studying the behavior of $T_{m}^{\ast }$
as a function of electron density, interaction strength etc. allows us to
obtain the phase diagram.

\section{Results}

We consider first small interaction strength $U=2.5t$ and $t^{\prime }=0.1t$%
. For this value of $t^{\prime }$ the ground state was previously found
unstable with respect to antiferromagnetic order and/or superconductivity at
the fillings close to van Hove band filling\cite%
{Metzner1,SalmHon,SalmHon1,KK}. Temperature dependences of the inverse
antiferromagnetic susceptibility ($\mathbf{Q}=(\pi ,\pi )$) obtained in the
present approach for different chemical potentials are shown in Fig. 1. One
can see that for large enough chemical potential $\mu >\mu _{c}^{\text{AF}%
}\approx -0.0375t$ ($\mu =0$ corresponds to van Hove band filling), the
inverse antiferromagnetic susceptibility monotonously decreases with
decreasing temperature and vanishes at a certain temperature $T_{\text{AF}%
}^{\ast }$. The value of $T_{\text{AF}}^{\ast }$ increases with increasing $%
\mu $.

Study of susceptibilities at the incommensurate wave vectors (see Fig. 2)
shows that close to $\mu _{c}^{\text{AF}}$ (in the range $-0.06t<\mu <-0.02t$%
) we have $T_{\text{\textbf{Q}}}^{\ast }>T_{\text{AF}}^{\ast }$ for some $%
\mathbf{Q}=(\pi ,\pi -\delta )$. Therefore an instability with respect to
incommensurate, rather than a commensurate magnetic order is expected in
this interval of $\mu $. At $\mu =\mu _{c}^{\mathbf{Q}}=-0.06t$ we obtain $%
T_{\text{\textbf{Q}}}^{\ast }=0$, which shows existence of a quantum
critical point below half filling. Near the quantum critical point we find $%
\chi _{\mathbf{Q}}^{-1}\sim T$, which is similar to the result of the
Hertz-Moriya-Millis theory\cite{HM,MS}.

The behavior of the inverse susceptibility with respect to the $d$-wave
superconducting order is shown in the inset of Fig. 2. Similarly to the
inverse antiferromagnetic susceptibility, it monotonously decreases upon
lowering temperature, with a different temperature dependence.

\begin{figure}[tb]
\includegraphics[width=8.7cm]{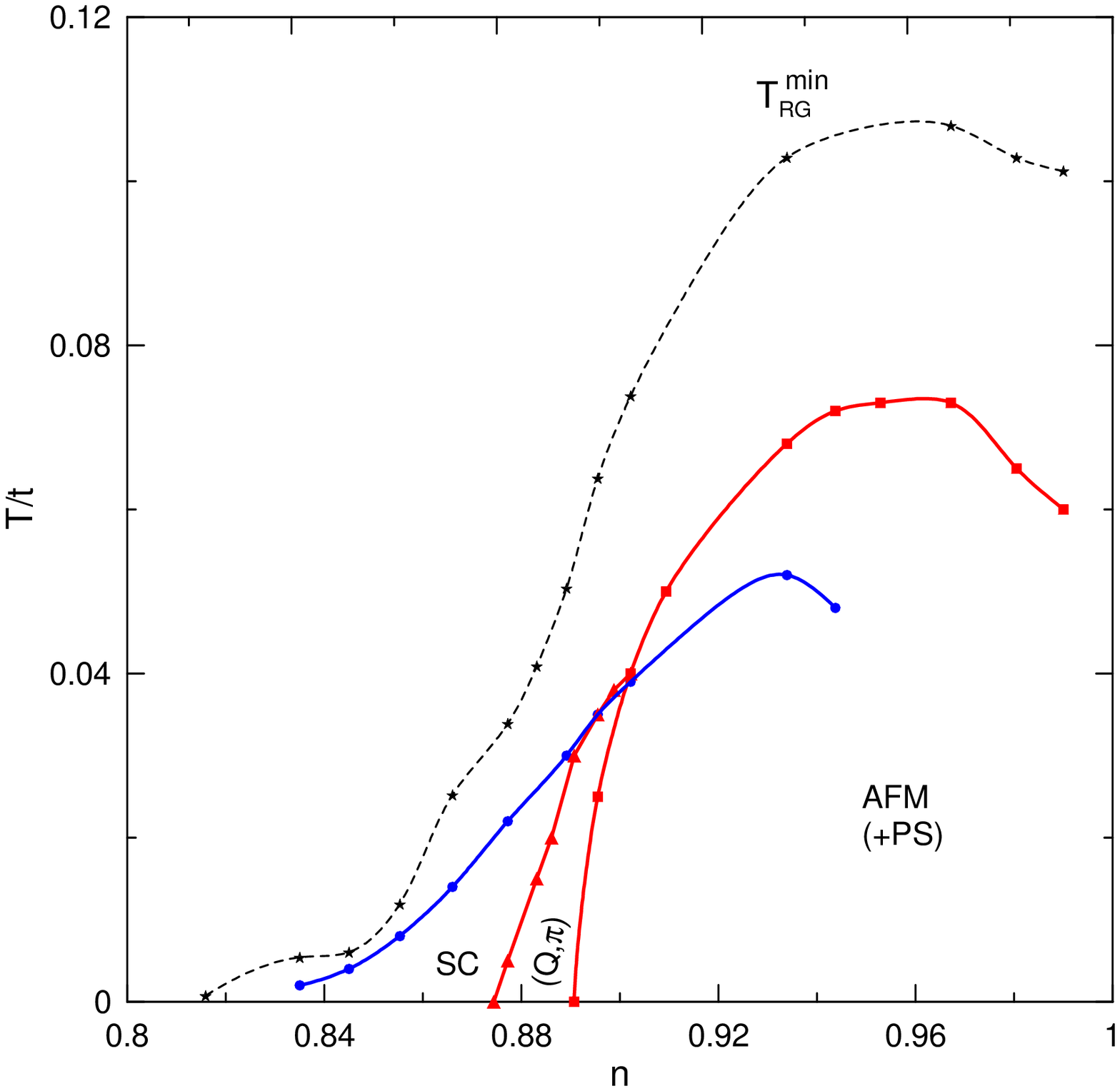} \vspace{-3mm}
\caption{(Color online) Phase diagram at $t^{\prime}/t=0.1$ and $U=2.5t$.
The temperatures of the crossover into regime with strong antiferromagnetic,
incommensurate magnetic, and superconducting fluctuations are marked by
squares, triangles and circles, respectively, PS denotes a possibility of
phase separation of the antiferromagnetic case. Dashed line (stars) show the
temperature $T^{\mathrm{min}}_{\mathrm{RG}}$, at which the fRG flow is
stoped }
\label{fig3}
\end{figure}

The obtained phase diagram is shown in Fig. 3 and contains
antiferromagnetic, incommensurate magnetic and superconducting phases. Away
from half filling the commensurate antiferromagnetic order is expected to be
unstable towards phase separation (to hole-rich and hole pure regions)\cite%
{PS}, although this possibility can not be verified in the present approach.
The obtained value of $T_{\text{dSC}}^{\ast }$ monotonously increases with
increasing density for $n\lesssim 0.94$. Deeper in the antiferromagnetic
phase the superconducting transition temperature is somewhat suppressed. The
origin of this suppression comes from the competition between
antiferromagnetic and superconducting fluctuations. The coexistence of
superconductivity and antiferromagnetism, which is possible in the interval $%
0.87<n<0.94,$ can not be verified in the present approach.

The region of the incommensurate phase obtained in Fig. 3 is much narrower,
than that expected in the mean-field approaches\cite{Arrigoni,Arzh}, which
predict incommensurate instability in the most part of the phase diagram. In
fact, accurate mean-field investigations\cite{Arrigoni,Yuki1,Petr} show,
that substantial part of incommensurate state in the mean-field approach is
unstable towards phase separation into commensurate and incommensurate
regions and therefore qualitatively agree with the renormalization group
approach. The presence of incommensurate phases within the renormalization
group approach was noticed previously for $t^{\prime}=0$ in Ref. \cite%
{Metzner1}.

The density dependence of $T_{\text{AFM}}^{\ast }(n)$ and $T_{\text{dSC}%
}^{\ast }(n)$, obtained in Fig. 3, is similar to that of the
antiferromagnetic and superconducting gap components in the electronic
spectrum, $\Delta _{\text{AFM}}(n)$ and $\Delta _{\text{dSC}}(n)$, recently
obtained within the combination of functional renormalization group approach
and mean-field theory\cite{Metzner}. Slower decrease of $T_{\text{dSC}%
}^{\ast }(n)$ when going into the antiferromagnetic phase in the present
approach is explained by the fact that in the present approach magnetic and
superconducting fluctuations are weaker coupled in the absence of
spontaneous symmetry breaking, since the latter leads to opening a gap in
the electronic spectrum at the Fermi surface. Contrary to the study of Ref. 
\cite{Metzner} we included incommensurate phases in our analysis.

\begin{figure}[tb]
\includegraphics[width=8.7cm]{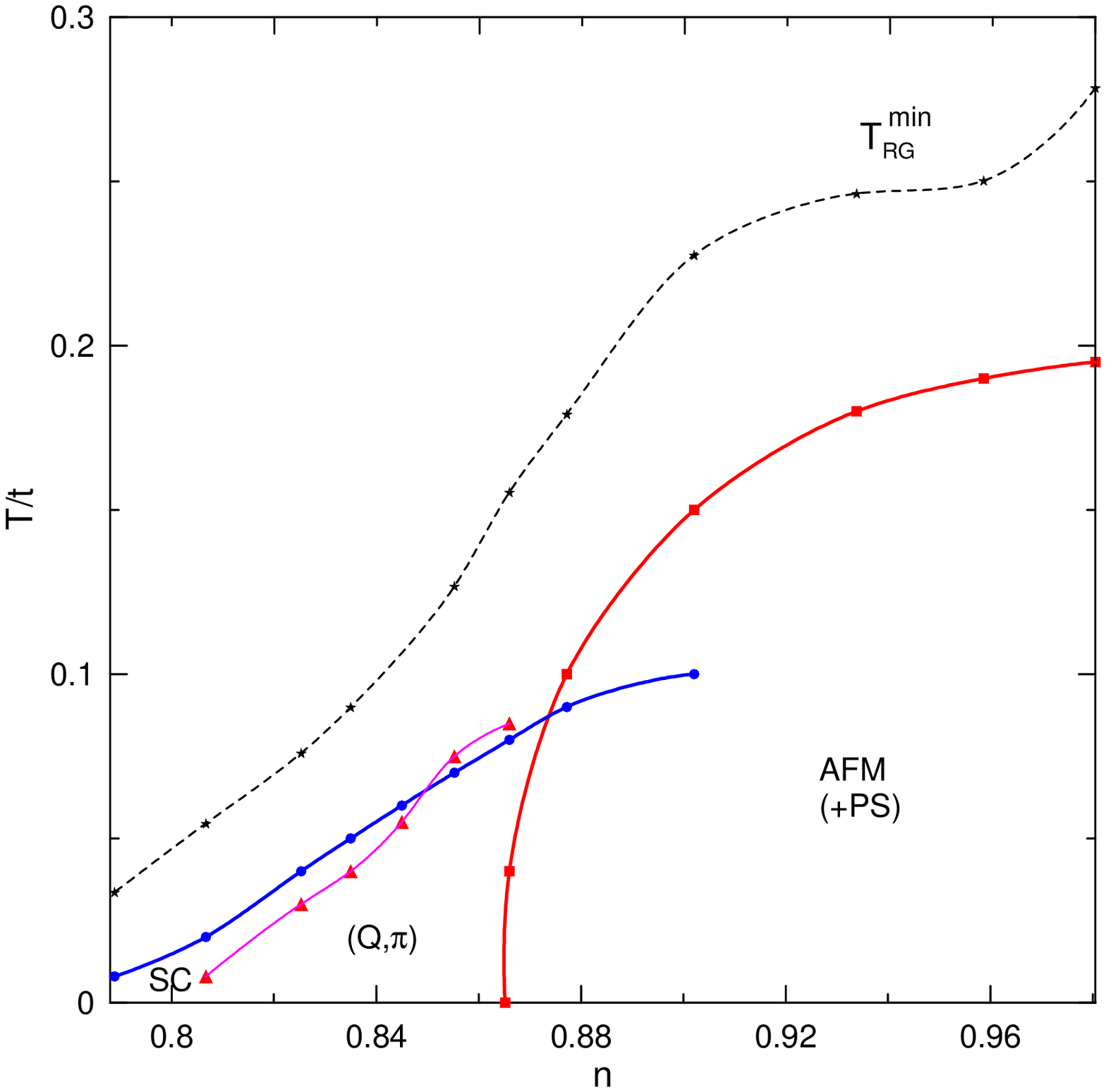} \vspace{-3mm}
\caption{(Color online) Phase diagram at $t^{\prime}/t=0.1$ and $U=3.5t$.
The notations are the same as in Fig. 3}
\label{fig4}
\end{figure}

\begin{figure}[tb]
\includegraphics[width=8.7cm]{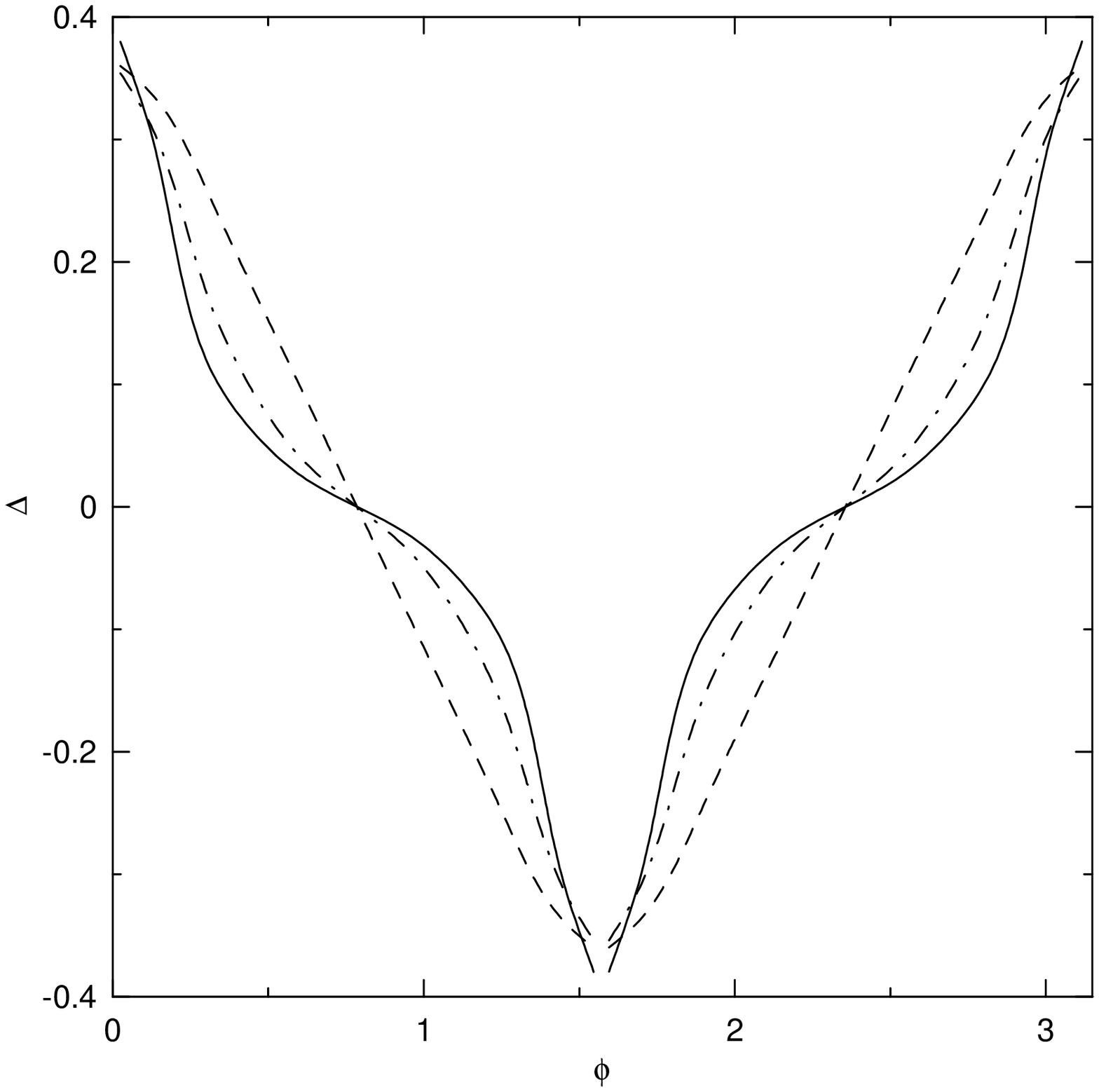} \vspace{-3mm}
\caption{Angular dependence of the superconducting gap for $U=2.5t$, $n=0.87$
(dot-dashed line) and $U=3.5t$, $n=0.84$ (solid line), $t^{\prime}/t=0.1$.
Dashed line shows the standard $\Delta=(\cos k_x-\cos k_y)/A$ dependence}
\label{fig5}
\end{figure}

At $U=3.5t$ we obtain similar behavior of the magnetic and superconducting
susceptibilities near the quantum critical point; the resulting phase
diagram is shown in Fig. 4. Compared to the case $U=2.5t$, the phase diagram
has broader region of the incommensurate phase. The crossover temperature
into regime with strong superconducting fluctuations approximately follows
that for the incommensurate fluctuations, implying that the
superconductivity in this case is possibly caused by incommensurate spin
fluctuations. To clarify this point, we plot in Fig. 5 the momentum
dependence of the superconducting gap, obtained from the Bethe-Salpeter
analysis\cite{BS}. We see that the shape of the gap, calculated for $U=3.5t$
shows stronger deviation from the $d$-wave form, than for $U=2.5t,$ which
indicates possible role of the incommensurate fluctuations in this case.

\section{Conclusion}

\bigskip We have investigated temperature dependence of the commensurate and
incommensurate magnetic susceptibilities, as well as the susceptibility with
respect to the $d$-wave pairing in the fRG framework, which allowed us to
obtain the phase diagrams of the Hubbard model at different $U.$ We obtain
an intermediate phase with strong incommensurate fluctuations between the
commensurate and paramagnetic phases, the former is characterized by a
wavevector $\mathbf{Q}=(\pi ,\pi -\delta ).$ The size of the incommensurate
phase increases with increasing interaction strength. The tendency towards
incommensurate order near magnetic quantum phase transition comes from the
absence of nesting of the Fermi surface at finite $t^{\prime }$. The
corresponding profile of static noninteracting spin susceptibility $\chi
_{0}(\mathbf{Q})$ is however almost flat near $\mathbf{Q}=(\pi ,\pi )$ (see,
e.g. Ref. \cite{Onufr}) showing that one can not restrict oneself to
fluctuations with only one certain $\mathbf{Q},$ as assumed in HMM theory.
At the same time, the obtained size of the incommensurate phase is much
narrower, than obtained in the mean-field approaches\cite{Arrigoni,Arzh},
which is explained by existence of a phase separation in both approaches.
Near the quantum critical point the inverse magnetic susceptibility with
respect to the preferable order parameter shows in fRG approach almost
linear temperature dependence, similar to that in HMM theory. The
electron-paramagnon interaction, not considered in the present study, may
however change the critical behavior of the susceptibility. Note that
recently the incommensurate magnetic fluctuations were also considered
within the renormalization-group approach for fermion-boson model\cite%
{Wetterich}, where similar results were obtained.

While the Mermin-Wagner theorem states no spontaneous breaking of continuos
symmetry in two dimensions at finite $T$, we have obtained finite
temperature of vanishing inverse magnetic and superconducting
susceptibilities, which is the consequence of the one-loop approximation,
considered in Eqs. (\ref{dV}). As we argue in the Introduction, the obtained
temperatures $T_{m}^{\ast }$ should be considered as a crossover temperature
to the regime with strong magnetic fluctuations and exponential increase of
the correlation length.

The patching scheme invoking the projection of the vertices to the Fermi
surface, used in the present renormalization group study, may have some
influence on the phase diagram. We expect, however, that this influence does
not modify the phase diagram strongly. This is confirmed by the recent
two-loop study \cite{TwoLoop} which necessarily includes corrections to the
effect of the projection of vertices and shows that the effects of these
corrections and the two-loop corrections to large extent cancel each other.

The non-analytical corrections to the susceptibility and electron-paramagnon
interaction vertices may become important near quantum phase transitions\cite%
{Chub}. These corrections are however expected to produce much weaker
effect, than the effects of the band dispersion considered in the present
paper. Investigation of the role of these corrections in the presence of van
Hove singularities has to be performed.

Application of the method considered in the present paper to ferromagnetic
instability and detail comparison of the results of the present approach
with the mean-field approach and quasistatic approach of Ref. \cite{Yuki1}
also has to be performed. 

\section{Acknowledgements}

I am grateful to H. Yamase for stimulating discussions and careful reading
of the manuscript. The work is supported by grants 07-02-01264a and
1941.2008.2 from Russian Basic Resarch Foundation and by the Partnership
program of the Max-Planck Society.

%\begin{figure}[tb]
%\psfig{file=Fig1.eps,width=90mm,silent=} \vspace{-3mm}
%\includegraphics[width=8.7cm]{Fig3a.eps} \vspace{-3mm}
%\caption{Phase diagram at $t^{\prime}/t=-0.1$ and $U=2.5t$}
%\label{fig1}
%\end{figure}

%\begin{figure}[tb]
%\psfig{file=Fig5.eps,width=90mm,silent=} \vspace{-3mm}
%\includegraphics[width=8.7cm]{Fig5.eps} \vspace{-3mm}
%\caption{Phase diagram at $t^{\prime}/t=-0.1$ and $U=2.5t$}
%\label{fig1}
%\end{figure}

\end{document}